\documentclass[twocolumn,prl]{revtex4}

\usepackage{amssymb,graphicx}

\newcommand{\cF}{\mathcal{F}}
\newcommand{\beq}{\begin{equation}}
\newcommand{\eeq}{\end{equation}}
\newcommand{\bea}{\begin{eqnarray}}
\newcommand{\eea}{\end{eqnarray}}

\begin{document}
\title{Measurement of energy eigenstates by a slow detector}
\author{I. Martin and W. H. Zurek}
\affiliation{Theoretical Division, Los Alamos National Laboratory, Los Alamos,
NM 87545, USA}
\date{\today}
\begin{abstract}

We propose a method for a weak continuous measurement of the energy eigenstates
of a fast quantum system by means of a ``slow" detector.  Such a detector is
only sensitive to slowly-changing variables, e. g. energy, while its
back-action can be limited solely to decoherence of the eigenstate
superpositions.  We apply this scheme to the problem of detection of quantum
jumps between energy eigenstates in a harmonic oscillator.

\end{abstract}

\pacs{...}

\maketitle

Observation of quantum mechanical behavior of macro-scale systems is one of the
most fascinating current problems in physics.  Several successes have been
claimed so far.  They include observation of avoided crossing between
counter-propagating current states in Josephson loops \cite{lukens}, coherent
evolution of macroscopic quantum states in charge \cite{nakamura} and phase
\cite{martinis} qubits, and interference experiments on very large molecules
\cite{zeilinger}.

Recently, experiments on fabricated nano-mechanical systems became feasible
\cite{roukes}, with the position measurement resolution approaching the
zero-point motion uncertainty \cite{schwab}.  However, the commonly used {\em
linear} position measurement has fundamental limitations due to the measurement
apparatus back-action, which ultimately drives the oscillator away from the
quantum regime \cite{caves,mozyrsky,clerk}.  Thus a question naturally arises
whether it is feasible to observe quantum behavior of nanomechanical systems by
means of a {\em non-linear} measurement procedure.  Specifically, since the
oscillator energy is quadratic in displacement $x$, coupling the measurement
apparatus to $x^2$ can naturally lead to the measurement and projection into
the energy eigenstates.  The main difficulty with this approach is that such
coupling requires very precise tuning \cite{santamore} in order to eliminate
the linear-$x$ coupling which would otherwise dominate at small displacements.

There is an intimate connection between the measurement process and interaction
with a general extrinsic environment since both represent coupling of a system
to a macroscopically large number of degrees of freedom.  The practical
difference is that the {\em classical} output of the measurement apparatus (or
``meter") is available to the observer, and thus one can ask questions, e.g.
about the signal to noise ratio, that is how much of the output can be
attributed to the system that one is measuring, and how much to the noise of
the apparatus \cite{bragisnky}.  Depending on the relative strengths of the
system's coupling to the environment and to the measurement apparatus, its own
behavior can be either predominantly affected by the environment ({\em
einselection}) \cite{zurek} or by the meter.  The latter case is typically
equivalent to coupling to a non-equilibrium environment, which nevertheless can
result in a crossover to effectively thermal behavior of the system with the
``temperature" determined by the degree of non-equilibrium in the meter (e.g.
bias voltage in a quantum point contact \cite{mozyrsky}).

When environment is weakly coupled to the system and is {\em slow} compared to
the intrinsic system dynamics (determined by its self-hamiltonian $H_0$), an
interesting possibility arises: While the environment can lead to the
decoherence of superpositions of the energy eigenstates, it can not effectively
cause transitions between them.  This is a consequence of the quantum adiabatic
theorem.  Decoherence is commonly associated with a gradual projection into one
of the states of the superposition. Thus one concludes that the system should
remain in the energy eigenstates \cite{paz} most of the time, with the rare
transitions (quantum jumps, QJ) between them governed by the residual
relaxation processes. A necessary condition for this scenario is a presence of
non-zero diagonal (in the energy eigen-basis) coupling between environment and
the system, which leads to environment-induced fluctuations in the system level
spacing.

In this Letter we apply the ideas of slow environment to the measurement
process and determine the circumstances under which QJ should become
observable. We perform analysis for a harmonic oscillator playing the role of
the quantum system being measured, with the Hamiltonian
$$H_o = \frac{p^2}{2m} + \frac {m\omega_0^2 x^2}{2},$$
where $x$ is the oscillator position, $p$ is the momentum, $m$ is its mass and
$\omega_0$ is its frequency.  Analysis for a more general systems is similar.
For harmonic oscillator, the requirement on non-zero diagonal coupling is
equivalent to inclusion of non-linear terms in the meter-oscillator coupling.
For concreteness we consider terms up to quadratic order in displacement,
\beq
H_{\rm int} = (\lambda_1 x + \lambda_2 x^2) \cF\equiv
\Omega(x)\cF.\label{eq:Hint}
\eeq
Note that for a slow operator $\cF$ the linear coupling term
($\lambda_1$) does not generate non-linear in $x$ terms within
perturbation theory.

We will first derive the measurability criteria for QJ for a
general measurement apparatus that interacts with the oscillator
via operator $\cF$ and generates output $I$. We will then give a
specific example of such a measurement when a single electron
transistor (SET) is capacitively coupled to the oscillator.  This
experimental procedure is closely related to the displacement
measurement techniques currently in use \cite{schwab}.

The complete Hamiltonian for the combined system of the oscillator and the
measurement apparatus (meter) consists of three parts,
\beq
H = H_o + H_M + H_{\rm int}
\eeq
with $H_o$ and $H_{\rm int}$ specified above and $H_M$ being the Hamiltonian of
the meter.  The evolution of the full density matrix $\rho$ is given by $\dot{
\rho} = i [\rho, H]$.  Assuming that the interaction between the measurement
apparatus and the system is relatively weak, that is it dominates neither the
oscillator's dynamics nor the meter's (``weak measurement"), we can determine
the evolution by means of the perturbation theory.  In the interaction
representation, where $\hat{\rho} = e^{-i(H_0+H_M)t}\rho e^{i(H_0+H_M)t}$ and
$\hat{H}_{\rm int} = e^{i(H_0+H_M)t} H_{\rm int} e^{-i(H_0+H_M)t}$, the
second-order perturbation theory yields,
\bea
\dot{\hat{\rho}} &=& i[\hat{\rho}, \hat{H}_{\rm int}]=
i{[\rho(0)+i\int_0^tdt'[\hat{\rho}, \hat{H}_{\rm int}]_{t'}, \hat{H}_{\rm int}]}\\
&\approx& i [\rho(0), H_{\rm int}] - \int_0^t{dt'[[\hat{\rho}, \hat{H}_{\rm
int}(t')], \hat{H}_{\rm int}]}
\eea
Here the operators with unspecified time variable are taken at time $t$.  The
final step is equivalent to the Markovian approximation, which is valid if the
time correlations in $\hat{H}_{\rm int}$ decay faster than the decay times of
diagonal and off-diagonal elements of the density matrix (in the jargon of
nuclear magnetic resonance, the times $T_1$ and $T_2$, respectively).  Making
now the common assumption that the initial density matrix is factorizable,
$\rho(0) = \rho_o(0)\bigotimes \rho_M(0)$, and tracing over the meter density
matrix $\rho_M$ we can obtain the evolution of the reduced density matrix of
the oscillator $\rho_o$.
\bea
\dot{\hat\rho}_o &=& i[\hat\rho_o, \hat\Omega]\langle\hat\cF\rangle \label{eq:rodot}\\
&-& \int_0^t dt'[[\hat\rho_o,\hat\Omega(t')], \hat\Omega]\frac{S_{\cF}(t,t')}{2} \nonumber\\
&+& \int_0^t
dt'[\{\hat\rho_o,\hat\Omega(t')\},\hat\Omega]\frac{A_{\cF}(t,t')}{2}\nonumber
\eea
The first term in Eq.~(\ref{eq:rodot}) corresponds to the renormalization of
the oscillator Hamiltonian by the interaction with the meter via the average
$\langle\cF (t)\rangle = Tr_M \hat\rho_M \hat{\cF}(t)$ and will be neglected in
the following discussion based on the assumption that the interaction between
the meter and the oscillator is week. The second and the third terms in
Eq.~(\ref{eq:rodot}) describe decoherence and relaxation of the oscillator
induced by the backaction of the meter.  Defining the noise correlator
$G_{\cF}(t,t') = \langle\cF(t)\cF(t')\rangle -
\langle\cF(t)\rangle\langle\cF(t')\rangle$, the symmetric and antisymmetric
correlators $S_{\cF}$ and $A_{\cF}$ are $G_{\cF}(t,t')\pm G_{\cF}(t',t)$. Note
that the antisymmetric correlator is only non-zero if the operator $\cF$ is
quantum, i.e. doesn't commute with itself at different times.

The density matrix evolution equation can be simplified by substituting
$\hat\Omega(t') = e^{iH_o(t'-t)}\hat\Omega(t)e^{-iH_o(t'-t)}$.  Assuming the
stationary noise correlator  $G(t,t') = G(t-t')$ which decays on the time scale
faster than $T_1$ and $T_2$, in the energy eigen-basis,
\bea
\dot{\hat\rho}_{nk} &=&
-\frac{\lambda_1^2\ell_0^2}{2}\left[{\tilde S_{\cF}}(\omega_0)(n+k+1)-{\tilde A_{\cF}}(\omega_0)\right]\,\hat\rho_{nk}\label{eq:rhonm}\\
&& +\lambda_1^2\ell_0^2\tilde
G_{\cF}(\omega_0)\sqrt{(n+1)(k+1)}{\hat\rho}_{n+1\,k+1}\nonumber\\
&& + \lambda_1^2\ell_0^2 \tilde G_{\cF}(-\omega_0)\sqrt{nk}
{\hat\rho}_{n-1\,k-1}\nonumber\\
&&-\frac{\lambda_2^2\ell_0^4}{4}{\tilde S_{\cF}}(0){(n-k)^2}\,\hat\rho_{nk}\nonumber\\
\dot{\hat\rho}_{nn} &=&
-{\lambda_1^2\ell_0^2}\left[{\tilde S_{\cF}}(\omega_0)(n+1/2)-{\tilde A_{\cF}}(\omega_0)\right]\,\hat\rho_{nn}\label{eq:rhonn}\\
&&+\lambda_1^2\ell_0^2{\tilde G_{\cF}}(\omega_0)(n+1)\,\hat\rho_{n+1,n+1} \nonumber\\
&&+ \lambda_1^2\ell_0^2{\tilde
G_{\cF}}(-\omega_0)n\,\hat\rho_{n-1,n-1}\nonumber
\eea
where $n\ne k$ in the first equation, and tilde denotes the Fourier transform.
The zero-point motion amplitude is $\ell_0 = \sqrt{\hbar/(2m\omega_0)}$. From
Eq.~(\ref{eq:rhonn}) the relaxation rate from state $n$ is
\beq
\frac{1}{T_{1n}} ={\lambda_1^2\ell_0^2}\left[{\tilde
S_{\cF}}(\omega_0)(n+1/2)-{\tilde A_{\cF}}(\omega_0)\right]\label{eq:T1m}
\eeq
We can similarly define from Eq.~(\ref{eq:rhonm}) the decoherence rate
corresponding to the decay of the off-diagonal density matrix element
$\hat\rho_{nm}$,
\beq
\frac{1}{T_{2nk}} =
\frac{1}{2T_{1n}}+\frac{1}{2T_{1k}}+\frac{\lambda_2^2\ell_0^4}{4}{\tilde
S_{\cF}}(0){(n-k)^2}.
\eeq
For slow measurement we expect that $T_{2nk} \ll T_{1n}$ for any $n\ne k$.

For the energy eigenstate $n$ to be measurable, the relaxation time for that
state, $T_{1n}$, has to be longer than the measurement time $T_M$.   Typically
$T_M$ is defined as the time needed to measure a variable $n$ with the
signal-to-noise ratio of 1. For low-frequency part of the detector output noise
$\tilde S_I(0)$, and the incremental change in the output $\Delta I = |I(n+1) -
I(n)|$, the measurement time is
\beq
T_M = \frac{\tilde S_I(0)}{(\Delta I)^2}
\eeq
From the linear response theory, we can define the forward $\Lambda =
(i/\hbar)\int_0^\infty{dt[\cF(0), I(t)]}$ and reverse $\Lambda' =
(i/\hbar)\int_0^\infty{dt[I(0), \cF(t)]}$ detector gains.  Then
\beq\label{eq:Iresp}
\Delta I = \lambda_2\ell_0^2 \Lambda\, \Delta n = \lambda_2\ell_0^2 \Lambda.
\eeq
The signal-to-noise ratio (SNR) achievable before the $n$-th state decays is
\bea
\eta_n &=& \frac{T_{1n}}{T_M} \approx
\frac{2}{(2n+1)}\frac{\lambda_2^2\ell_0^2}{{\lambda_1^2}}\frac{\Lambda^2}{\tilde
S_I(0) {\tilde S_{\cF}}(\omega_0)}\\
&\le&\frac{8}{(2n+1)}\times\frac{\lambda_2^2\ell_0^2}{{\lambda_1^2}}\times\frac{S_{\cF}(0)
}{S_{\cF}(\omega_0)}.\label{eq:snr}
\eea
To derive the last inequality we used the Schwartz inequality, $S_I S_\cF \le
|S_{I\cF}|^2$, the identity $\Lambda - \Lambda' = (2/\hbar)\, {\rm Im}
S_{I\cF}$, and the absence of positive feedback in the detector, $\Lambda \cdot
\Lambda' <0$ \cite{clerk}.  The conditions for the equality in
Eq.~(\ref{eq:snr}) include absence of reverse gain, $\Lambda' = 0$, and ${\rm
Re} S_{I\cF} = 0$.  For realistic nanomechanical systems' frequencies and
temperatures, the first factor can be of order 1, the second is likely to be
small due to the square of the zero-point-motion amplitude, and the last one,
by construction, for the slow detector can be very large.  Thus as a matter of
principle the measurement of the number states by a slow detector is possible.

In addition to the coupling to the meter, an oscillator is influenced by its
own environment, which leads to a finite intrinsic quality factor $Q$.  The
contribution of this heat bath to relaxation at $T > \hbar \omega_0$ is
\beq
\frac{1}{T_{1n}^Q} = \frac{T}{Q} n,\label{eq:T1env}
\eeq
which may reduce the SNR of Eq.~(\ref{eq:snr}).

\begin{figure}[htb]

\vspace{-0 mm} \centerline{\includegraphics[width=2.8 in]{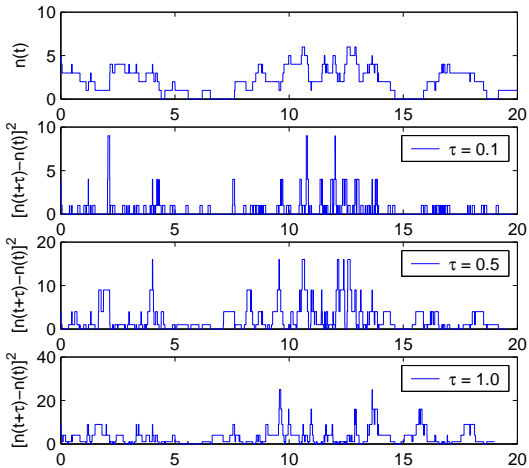}}
\caption{Simulation of the oscillator occupation number dynamics, described by
Eq.~(\ref{eq:rhonn}).  The parameters of simulation are $\Gamma_{up} = 1$,
$\Gamma_{dn} = 1.2$ } \label{Fig:nt}\vspace{-0 mm}
\end{figure}

The output of a detector with high SNR, $\eta_n\gg 1$, is proportional to
$n(t)$ (QJ's are not masked by the detector noise).  Stochastic simulation of
this process is shown in the top panel of Fig.~\ref{Fig:nt}.  The switching
between the plateaus corresponding to different values of $n$ occurs with the
average rate $\bar\Gamma\sim(\Gamma_{up}+\Gamma_{dn})\bar n =\Gamma_{up}
(\Gamma_{up}+\Gamma_{dn})/(\Gamma_{dn}-\Gamma_{up})$, where
$\Gamma_{up}=\lambda_1^2\ell_0^2{\tilde G_{\cF}}(\omega_0)$ is the transition
rate from the ground state into the first excited ($0\rightarrow 1$), and
$\Gamma_{dn}=\lambda_1^2\ell_0^2{\tilde G_{\cF}}(-\omega_0)$ is the opposite
rate ($1\rightarrow 0$), see Eq.~(\ref{eq:rhonn}).  The fast switching is
superimposed on top of the slow dynamics occurring on the time scale of the
inverse damping coefficient, $\gamma^{-1} = (\Gamma_{dn} - \Gamma_{up})^{-1}$.

In the presence of a detector noise the QJ may not be directly observable in
$I(t)$.  However, there still may be signatures in the signal correlations.
First let's consider the two-point time autocorrelation function $C_2(t_1,t_2)
= \langle n(t_1) n(t_2)\rangle$ which enters the output autocorrelation
function, $\langle I(t_1) I(t_2)\rangle$.  It can be expressed in terms of the
stationary probability distribution $P(n)$ and the conditional probability that
the system is in state $n_2$ at time $t_2$ given that it was in state $n_1$ at
an earlier time $t_1$, $P(n_2,t_2;n_1,t_1)$, as $C_2(t_1,t_2) =
\sum_{n_1,n_2}{n_2 P(n_2,t_2;n_1,t_1) n_1 P(n_1)}$.   From the equation of
motion for $\tilde n(t) = \sum_n{n \rho_{nn}(t)}$ that can be obtained from
Eq.~(\ref{eq:rhonn}), it is easy to see, however, that $C_2$ is not sensitive
to the fast switching with rate $\bar\Gamma$.
\begin{figure}[htb]

\vspace{-0 mm} \centerline{\includegraphics[width=2.8 in]{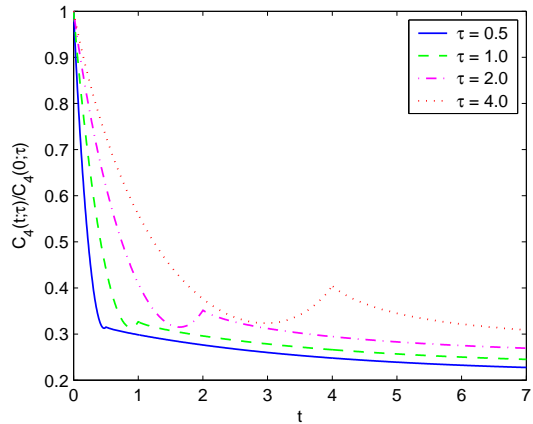}}
\caption{Four-point correlation function containing the information about fast
switching between the number states. The parameters of simulation are
$\Gamma_{up} = 1$, $\Gamma_{dn} = 1.2$ } \label{Fig:c4}\vspace{-0 mm}
\end{figure}

Therefore, in order to extract the signatures of the fast switching, we  need
to go to higher-order correlators.  One approach is to convert the multi-level
staircase dynamics into a two-level, or {\em telegraph} switching. This is
achieved by performing the following non-linear transformation on the output
signal: $I(t) \rightarrow [I(t+\tau) - I(t)]^2$.  It will be telegraph-like if
$\bar \Gamma \tau \ll 1$.  Examples for different values of $\tau$ are shown in
the lower three panels of Fig.~\ref{Fig:nt}.  We now define the four-point
correlation function,
$$C_4(t_1,t_2;\tau) = \langle[n(t_2+\tau)- n(t_2)]^2[n(t_1+\tau) - n(t_1)]^2\rangle,$$
which is expected to be sensitive to the fast switching.  The contributions to
this correlation have the general form $\langle
n(t_4)n(t_3)n(t_2)n(t_1)\rangle$, with $t_4>t_3>t_2>t_1$. Similar to the
two-point correlation function, it can be expressed as
\begin{eqnarray}
\sum_{n_1,n_2,n_3,n_4}n_4 P(n_4,t_4;n_3,t_3) n_3 P(n_3,t_3;n_2,t_2)\\
\times n_2 P(n_2,t_2;n_1,t_1) n_1 P(n_1).\nonumber
\end{eqnarray}
Rewriting Eq.~(\ref{eq:rhonn}) in the matrix form, $\dot\rho = M\rho$, the
transition matrices can be expressed as
\begin{eqnarray}
P(n_2,t_2;n_1,t_1) = U \exp[D(t_2-t_1)] U^{-1},
\end{eqnarray}
where $D$ is a diagonal matrix of eigenvalues and $U$ is the matrix of
eigenvectors of $M$, $MU = U D$.  Performing numerical diagonalization on the
truncated system of equations (\ref{eq:rhonn}), we obtain the correlator
$C_4(t;\tau) \equiv C_4(t,0;\tau)$ for various values of $\tau$, as shown in
Fig.~\ref{Fig:c4}. Indeed we find a qualitative change when the delay time
$\tau$ becomes comparable to the average switching time, $\bar \Gamma \tau
\approx 1$. For short delay times, $\bar \Gamma \tau \ll 1$, the correlation
function rapidly decays at short times ($t<\tau$), followed by a slow
exponential decay at long times. It is easy to see that in this regime, the
correlation function has to be linear, $C_4 \propto \tau - t$ for $t\le \tau$.
For long delay times, $\bar \Gamma \tau > 1$, the correlation function is a sum
of exponentials.  When an additive detector noise $\xi$ with short correlation
time $\tau_\xi << {\bar \Gamma}^{-1}$ is included, the behavior of $C_4$ only
changes at $t<\tau_\xi$.  Thus, as far as $\tau_\xi < \tau$, the extrinsic
noise can be effectively filtered out and $C_4$ can be used to detect QJ's.

\begin{figure}[htb]
\vspace{-0 mm} \centerline{\includegraphics[width=2.2 in]{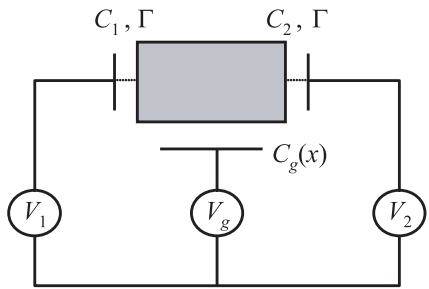}}
\caption{Scheme for SET displacement detector.  Position of the oscillator
modulates capacitance $C_g(x)$, thus modifying the current between the left and
right leads.}\label{Fig:SET} \vspace{-0 mm}
\end{figure}

A particular example of a measurement apparatus that can be used to perform the
slow measurement described above is a Single Electron Transistor (SET).  The
nanoresonator interacts with the SET island via displacement-dependent gate
capacitance $C_g(x)=C_g^0/(1-x/d)$, where $d$ is the distance between the
nanoresonator and the SET island (Fig.~\ref{Fig:SET}). The displacement is read
out by measuring the current $I$ through the SET.  The island is also coupled
to the leads via capacitances $C_1$ and $C_2$.  For concreteness assuming that
$C_g^0 = C_1+C_2$, the interaction Hamiltonian is
\beq
H_{\rm int} = \frac12 qV_g\left(\frac{x}{2d} + \frac{x^2}{4d^2}\right),
\eeq
which indeed has the form of Eq.~(\ref{eq:Hint}), with $\cF = q$ the island
charge, $\lambda_1 = V_g/(4d)$ and $\lambda_2 = V_g/(8d^2)$.  To satisfy the
conditions of ``slow" measurement, the SET tunneling rate has to be much slower
than the oscillator frequency, $\Gamma \ll \omega_0$.  We consider the
``threshold" regime, which corresponds to the resonant level nearly aligned
with one of the leads' chemical potential and corresponds to the highest
read-out sensitivity. We assume that either $T\ll \Gamma$ or that SET is
superconducting.  Then,
\beq
S_q(0) \approx \frac{1}{\Gamma},\quad S_q(\omega_0) \approx
\frac{\Gamma}{\omega_0^2}.
\eeq
SET is a nearly ideal detector \cite{set}; thus, the conditions for equality in
Eq.~(\ref{eq:snr}) are approximately satisfied.  Assuming that the intrinsic
(measurement-unrelated) relaxation of the oscillator is smaller than the
measurement-induced one, the signal-to-noise ratio is
\bea
 \eta_n^{SET}
 \sim\frac{8}{(2n+1)}\times\frac{\ell_0^2}{4d^2}
  \times\frac{\omega_0^2}{\Gamma^2}\label{eq:snr2}
\eea
From Eqs.~(\ref{eq:T1m}) and (\ref{eq:T1env}) the measurement-induced
relaxation will dominate the environmental one if
$\lambda_1^2\ell_0^2Q/(T\Gamma) > 1$.

For a numerical estimate we take an oscillator with the frequency $\omega_0 =
2\pi\times100$ MHz and the mass $m = 2.5\cdot10^{-17}$ (for 1$\ \mu{\rm
m}\times 0.1\ \mu{\rm m}\times 0.1\ \mu{\rm m}$ dimensions at the Si density
2,500 kg/m$^3$).  The zero-point-motion amplitude is $\ell_0 = 10^{-13}$ m.
Assuming that the effective separation between the nanoresonator and the SET
island is $d = 1$ nm and $\Gamma = 10^4\ {\rm s^{-1}}$, for SNR we find
$\eta_n^{SET}\sim 100/(2n+1)$.  It is easy to check that the intrinsic
relaxation (finite $Q$) under these conditions is insignificant compared to the
SET-induced relaxation.  At temperature $T = 10$ mK (superconducting SET) the
thermal occupation number of the resonator is about 2.  Therefore, under these
conditions one can expect to observe QJ between values of current that
correspond to different energy eigenstates of the nanoresonator.

\acknowledgments We would like to thank D. Mozyrsky and A. Shnirman for useful
discussions. This work was supported by the U. S. Department of Energy Office
of Science through contracts No. W-7405-ENG-36.

\end{document}